\documentclass[aps,showpacs,twocolumn]{revtex4}
\usepackage{amsfonts}
\usepackage{txfonts}
\usepackage{amssymb}
\usepackage{graphicx}
\usepackage{leftidx}

\newcommand{\bbra}{\{}
\newcommand{\kket}{\}}

\begin{document}
\title{Realistic Interpretation of Quantum Mechanics and Encounter-Delayed-Choice Experiment}
\author{ Gui-Lu Long$^{1,2,3}$, Wei Qin$^1$, Zhe Yang$^1$ and Jun-Lin Li$^1$}
\affiliation{$^1$ State Key Laboratory of Low-Dimensional Quantum Physics and Department of Physics, Tsinghua University, Beijing 100084,
China\\
$^2$Innovative Center of Quantum Matter, Beijing 100084, China\\
 $^3$ Tsinghua National Laboratory for Information Science and Technology, Tsinghua
University, Beijing 100084, China\\}
\date{\today}
\begin{abstract}
A realistic interpretation(REIN) of wave function in quantum mechanics is briefly presented in this work.
 In REIN, the wave function of a microscopic object is just its real existence rather than a mere mathematical description. Quantum object can exist in disjoint regions of space which just as the wave function distributes,  travels at a finite speed, and collapses instantly upon a measurement.
The single photon interference in a Mach-Zehnder interferometer is analyzed using REIN. In particular, we proposed
and experimentally implemented a generalized delayed-choice experiment, the encounter-delayed-choice(EDC) experiment, in which the second beam splitter is inserted at the encounter of the two sub-waves from the two arms. In the EDC experiment, the front parts of wave functions before the beam splitter insertion do not interfere and  show the particle nature, and the back parts of the wave functions will interfere and show a wave nature. The predicted phenomenon is clearly demonstrated in the experiment, and supports the REIN idea.
\end{abstract}
\maketitle
\date{\today}

The wave-particle duality is a central concept of quantum mechanics and is strikingly illustrated
in the well-known Wheeler's delayed-choice gedanken experiment \cite{VW1, VW2, Marlow, Hellmut, Lawson,Kim, Jacques1, Jacques2, Ma}. A good demonstration of the
delayed-choice experiment is given by a two-path interferometer, Mach-Zehnder interferometer (MZI),
seen in Figure \ref{WDC}(a). A single photon is directed to the MZI followed by two detectors at its end.
If the output beam splitter BS$_{2}$ is present (closed configuration), the photon is first
split by the input beam splitter BS$_{1}$ and then travels inside the MZI with a tunable
phase shifter $\phi$ until the two interfering paths are recombined by BS$_{2}$. When $\phi$ is varied,
the interference fringes are observed as a modulation of the detection probabilities of detectors
D$_{1}$ and D$_{2}$. It indicates that the photon travels both
paths of the MZI to behave as a wave and the two paths are indistinguishable. If BS$_{2}$ is absent
(open configuration), a click in only one of the two detectors with probability $1/2$, independent of $\phi$,
is associated with a given path to indicate that the photon travels along a single path and behaves
as a particle. Such an experiment concludes that quantum systems exhibit
wave or particle behavior depending on the configuration of the measurement apparatus. Moreover, the
two complementary experimental setups are mutually exclusive and the two behaviors, wave and particle
behavior, cannot be observed simultaneously.

Recently, a new extension of the delayed-choice experiment (quantum delayed-choice) \cite{Ionicioiu,
Schirber, Roy, Auccause, Peruzzo, Kaiser, GGC, Adesso}, where
the output beam splitter in this classical state is replaced with that in a quantum
superposition state, has been proposed. The experiment indicates that BS$_{2}$ can be simultaneously
absent and present, and both wave and particle behavior can be simultaneously observed
to show a morphing behavior between wave and particle.

The concept of a wave function is introduced to quantum theory as a completely description
of a quantum system. Wave function usually can be determined through tomographic methods, and be
measured directly by sequential measurements of two complementary variables relying on the
weak measurement \cite{Lundeen, kocsis, Schleich}.
It is the heart of quantum theory and its typical
interpretation is provided
by the Copenhagen interpretation \cite{Landau}, in which the wave function is treated as a complex probability amplitude in a pure mathematical manner. The essential understanding of the wave function has not been solved yet so far \cite{cohen,Mermin}.

In this article, we propose a realistic interpretation, the REIN,  on the wave function in quantum mechanics. Then we propose a generalized delayed-choice experiment, the encounter-delayed-choice (EDC) experiment to test the REIN. The EDC is experimentally demonstrated, and the results agree with the theoretical interpretation very well, which supports the idea of REIN. In the following, we will first present the main points of REIN. Then we describe the EDC experiment proposal. The experimental demonstration of the EDC proposal is followed. Finally we give a discussion and summary.

\vspace{0.5cm}
{\noindent \bf Results}\ \

{\bf The REIN.}
The essential idea of REIN is that wave function is realistic existence rather than just mathematical description. Here we give a brief introduction, and a detailed description will be given elsewhere \cite{rein}.

Quantum object, an object that obeys quantum mechanics, exists in the form of its wave function: extended in space and even in disjoint regions of space in some case. It changes forms as the wave function changes frequently. Since a wave function is usually a complex function, it has both amplitude and phase. If we just look at its spatial distribution, the square of the modulus of the wave function gives this distribution. However, it also has phase, and when two sub-wave functions merge or encounter, the resulting wave function will change differently at different locations: some is strengthened due to constructive interference whereas some other is canceled due to destructive interference. Thus a photon in a MZI is an extended object that exists in both arms. In the REIN view, there is no difference for a photon in a closed MZI setting and that in an open setting before they arrive at the second beam splitter. It also easier to comprehend how a photon can travel both arms. In REIN, a photon is an extended and  separated objects that exists simultaneously at both arms, just like a segment stream of water is divided into two branches, each then flows on its own in its riverbed. Of course, the quantum wave function is more powerful than the water stream as it has also a phase factor that gives rise to interference when it encounters with other sub-wave functions.

A sub-wave function is part of the whole wave function, for instance, the wave function in the upper arm of MZI,
needs not be normalized \cite{duality}. To emphasize, we use
 $|\psi\kket$ and $\bbra \psi|$ to denote a sub-wave function throughout this article.

The extended quantum wave function, the true or realistic quantum object, moves at a speed less or equals to the speed of light. As we know, light, an ensemble of photons, takes time to travel from the Sun to our planet. The electrons in a cyclotron travels slower than the light.

Quantum wave function, or quantum object, can change form by transformation or by measurement. It is easy to visualize the change in the wave function, but is difficult to visualize the change in a quantum object. This difficulty is pertinent to our stubborn notion of a rigid particle for a microscopic object, as the name quantum particle suggests. If we adopt the view that the quantum object does exist in the form of the wave function, it will be very easy to understand this change in form. Hence a photon wave function changes into two sub-wave functions when it is transformed by a beam splitter.

A measurement changes the shape, or form of a quantum object drastically.  According to the measurement postulate of quantum mechanics, a measurement will collapse the wave function instantly into one of the eigenstate of the measured observable. This change of the quantum object takes no time, and it is within all the spaces occupied by the wave function, which are disjoint in some cases. The measurement postulate cannot be derived from the Schroedinger equation, which governs the evolution of the quantum wave function. At this stage, one should not ask why measurement has such dramatic effect. The quantum object behaves just in this way. It is Nature.

\vspace{0.5cm}
{\bf EDC Experiment Proposal}
According to REIN, a photon is considered as the whole spatial distribution of its wave function, which really exists, more than a mere mathematical description. A new interpretation
of the single photon interference experiment in the MZI is given in the point of view of REIN.
The action of a $50/50$ beam splitter can be described by a so-called Hadamard transformation
given by
\begin{equation}\label{hadamard}
H=\frac{1}{\sqrt{2}}
\left(
\begin{array}{cc}
    1 & 1  \\
    1 & -1 \\
\end{array}
\right).
\end{equation}
When a single photon with its wave function $|\psi\rangle_{i}$ is directed to the MZI,
BS$_{1}$ works as a divider to split the wave function to two sub wave functions, $|\psi\}
_{in,1}$ and $|\psi\}_{in,2}$, traveling along path$_{1}$ and path$_{2}$ as
\begin{equation}
\left(
\begin{array}{c}
   |\psi\}_{in,1} \\
   |\psi\}_{in,2}  \\
\end{array}
\right)=H\left(
\begin{array}{c}
   |\psi\rangle_{i} \\
   0  \\
\end{array}
\right),
\end{equation}
which gives that $|\psi\}_{in,1}=|\psi\}_{in,2}
=|\psi\rangle_{i}/\sqrt{2}$. After a phase shifter $\phi$, an additional phase $e^{i\phi}$
is introduced and $|\psi\}_{in,1}$ becomes $e^{i\phi}|\psi\}_{in,1}$.
If BS$_{2}$ is absent, the two sub wave functions
are directed to the two detectors D$_{1}$ and D$_{2}$ without interference between
them. The detection probabilities of D$_{1}$ and D$_{2}$ are
$P_{1}=\leftidx{_{in,1}}\{\psi|\psi\}_{in,1}=1/2$ and
$P_{2}=\leftidx{_{in,2}}\{\psi|\psi\}_{in,2}=1/2$.
The sub-waves exist at both arms. There is equal probability the photon to collapse in either detectors. When a click is registered in D$_{1}$ (D$_{2}$), both of the two sub-wave functions collapse
to D$_{1}$ (D$_{2}$) instantly. In standard interpretation, this open MZI is usually interpreted as showing the particle nature. In contrast, the REIN interprets it still as realistic quantum waves. The two sub-waves from the two arms do not encounter, and both of them arrive at the two detectors. According to the measurement postulate of quantum mechanics, the measurement result will be one of the eigenstates, the eigenstates of discrete positions at D$_1$ and D$_2$,  with some probability.

If BS$_{2}$ is present, the coalescence  of the two sub-waves  occurs
to form two new sub-waves $|\psi\}_{out,1}$ and $|\psi\}_{out,2}$, which are
directed to D$_{1}$ and D$_{2}$ respectively. After the transformation of BS$_{2}$, we have
\begin{equation}
|\psi\}_{out,1}
=\frac{1}{\sqrt{2}}(e^{i\phi}|\psi\}_{in,1}-|\psi\}_{in,2})
\end{equation}
and
\begin{equation}
|\psi\}_{out,2}
=\frac{1}{\sqrt{2}}(e^{i\phi}|\psi\}_{in,1}+|\psi\}_{in,2}).
\end{equation}
The detection probabilities of D$_{1}$ and D$_{2}$ are
$P_{1}=\leftidx{_{out,1}}\{\psi|\psi\}_{out,1}=\sin^{2}\frac{\phi}{2}$
and
$P_{2}=\leftidx{_{out,2}}\{\psi|\psi\}_{out,2}=\cos^{2}\frac{\phi}{2}$. As $\phi$ varies, an interference pattern will appear. This has been used to show wave behavior in a closed MZI setting experiment. However, in the point view of REIN,
the quantum wave behaves exactly the same as that in the open MZI before reaching the end of the MZI. The insertion of BS$_2$ make the two sub-waves encounter and interfere due to their phases.
 Like in the open MZI, when a click is registered in D$_{1}$ (D$_{2}$), both of the
two output sub-waves collapse to D$_{1}$ (D$_{2}$) simultaneously.
In the special case where $\phi=0$, $|\psi\}_{in,1}$ and $|\psi\}_{in,2}$
interfere constructively to give that $|\psi\}_{out,2}=|\psi\rangle_{i}$ along
path$_{2}$, and interfere destructively to give $|\psi\}_{out,1}=0$ along path$_{1}$.
In this case, only D$_{2}$ can detect the photon.

\begin{widetext}
\begin{center}
\begin{figure}[!ht]
\includegraphics[width=16cm,angle=0]{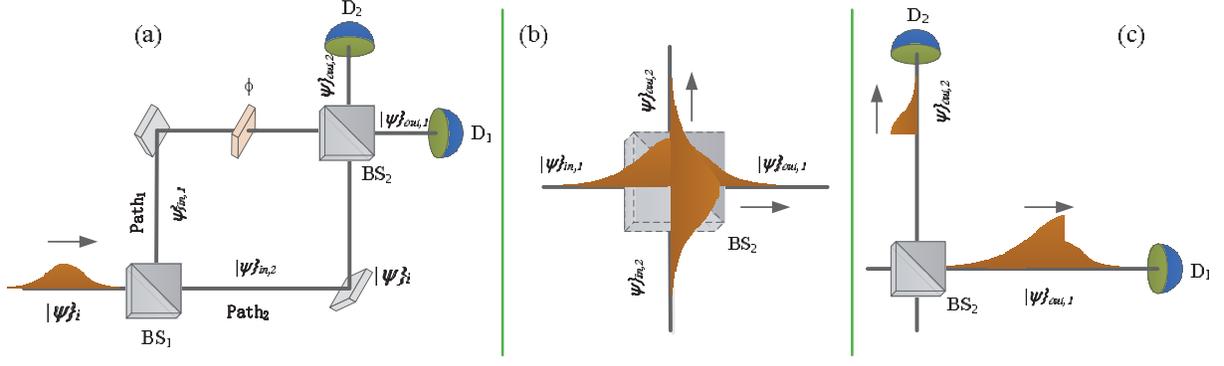}
\caption{(a) A Mach-Zehnder interferometer (MZI) with a tunable phase $\phi$ between its two arms. In the delay-choice MZI, the decision whether or not to insert BS$_2$ is made after the photon has reached the MZI, but has not arrived at the intended position of BS$_2$ (the exit point); (b) In the encounter-delayed-choice experiment, the insertion of BS$_{2}$
is made right at the encounter of the two sub-waves. As shown here, the front parts of the sub-waves have passed the exit point, while the back parts of the sub-waves have not passed through the exit point and are "closed" by BS$_2$;
(c) Still in EDC experiment, the two sub-waves leave the MZI and continue to move forward to D$_1$ and D$_2$. The front parts of the sub-waves retain their shape before they leave the MZI, but the back parts of the sub-waves are changed by the inserted BS$_2$. The back part of the up-going sub-wave vanishes due to destructive interference, whereas the right-going part of the sub-wave increases due to the constructive interference due to BS$_2$. The interference pattern of back parts of the sub-waves may vary according to their relative phases. }\label{WDC}
\end{figure}
\end{center}
\end{widetext}
If it is decided to insert BS$_{2}$ at the end of the MZI
when the two sub-waves encounter at the end of the MZI, $|\psi\}_{in,\rho}$
can be divided into two components and expressed as
\begin{equation}
|\psi\}_{in,\rho}=|\psi\}_{in,\rho}^{p}+|\psi\}_{in,\rho}^{w},
\end{equation}
with $\rho=1,2$. Here, $|\psi\}_{in,\rho}^{p}$ is the part of the sub-wave which has passed
the exit point while BS$_{2}$ has not been inserted, and they do not pass BS$_2$. $|\psi\}_{in,\rho}^{w}$ is the part of the sub-wave which has passed
the exit point while BS$_{2}$ has been inserted, and they will be subject to the action of BS$_2$. The interference between $|\psi\rangle_{in,1}^{w}$ and $|\psi\rangle_{in,2}^{w}$
occurs because BS$_{2}$ is present when they leave MZI. After the second beamsplitter,  it gives
\begin{equation}\label{outp1}
|\psi\}_{out,1}^{w}=\frac{1}{\sqrt{2}}
(e^{i\phi}|\psi\}_{in,1}^{w}-|\psi\}_{in,2}^{w})
\end{equation}
and
\begin{equation}\label{outp2}
|\psi\}_{out,2}^{w}=\frac{1}{\sqrt{2}}
(e^{i\phi}|\psi\}_{in,1}^{w}+|\psi\}_{in,2}^{w}),
\end{equation}
where $|\psi\}_{out,\rho}^{w}$ is the component of
$|\psi\}_{out,\rho}$ which will give the wave behavior in standard interpretation.
The interference
between $|\psi\}_{in,1}^{p}$ and $|\psi\}_{in,2}^{p}$ never occurs because BS$_{2}$
is absent when they exit out of the MZI. They are directed to the detectors along their paths
and we have
\begin{equation}\label{outw1}
|\psi\}_{out,1}^{p}=e^{i\phi}|\psi\}_{in,1}^{p}
\end{equation}
and
\begin{equation}\label{outw2}
|\psi\}_{out,2}^{p}=|\psi\}_{in,2}^{p},
\end{equation}
where
$|\psi\}_{out,\rho}^{p}$ is the component of $|\psi\}_{out,\rho}$ that will give
the particle behavior in standard interpretation. Combining equations (\ref{outp1}), (\ref{outp2}),
(\ref{outw1}) and (\ref{outw2}), we have the two new sub-waves after the action of BS$_{2}$
\begin{equation}
|\psi\}_{out,1}=|\psi\}_{out,1}^{p}+\frac{1}{\sqrt{2}}
(e^{i\phi}|\psi\}_{in,1}^{w}-|\psi\}_{in,2}^{w})
\end{equation}
and
\begin{equation}
|\psi\}_{out,2}=|\psi\}_{out,2}^{p}+\frac{1}{\sqrt{2}}
(e^{i\phi}|\psi\}_{in,1}^{w}+|\psi\}_{in,2}^{w})
\end{equation}
Ensuring the two paths inside the MZI are of equal length, we have
$|\psi\}_{in,1}^{p}=|\psi\}_{in,2}^{p}$ and
$|\psi\}_{in,1}^{w}=|\psi\}_{in,2}^{w}$.
The detection probabilities of D$_{1}$ and D$_{2}$ are
\begin{equation}
P_{1}=2\sin^{2}\frac{\phi}{2}P_{1}^{w}+P_{1}^{p},
\end{equation}
and
\begin{equation}
P_{2}=2\cos^{2}\frac{\phi}{2}P_{2}^{w}+P_{2}^{p},
\end{equation}
respectively. Here the relation
\begin{equation}
\leftidx{_{in,\rho}^{p}}\{\psi|\psi\}_{in,\rho}^{w}=0
\end{equation}
is employed,  and $P_{\rho}^{w}=\leftidx{_{in,\rho}^{w}}\{\psi|\psi\}_{in,\rho}^{w}$
($P_{\rho}^{p}=\leftidx{_{in,\rho}^{p}}\{\psi|\psi\}_{in,\rho}^{p}$)
is the probability that will (will not) show interference  behavior in the $\rho$-th arm.
 They satisfy the relation
\begin{equation}
P_{\rho}^{p}+P_{\rho}^{w}=\frac{1}{2}.
\end{equation}
\begin{center}
\begin{figure}[!ht]
\includegraphics[width=8cm,angle=0]{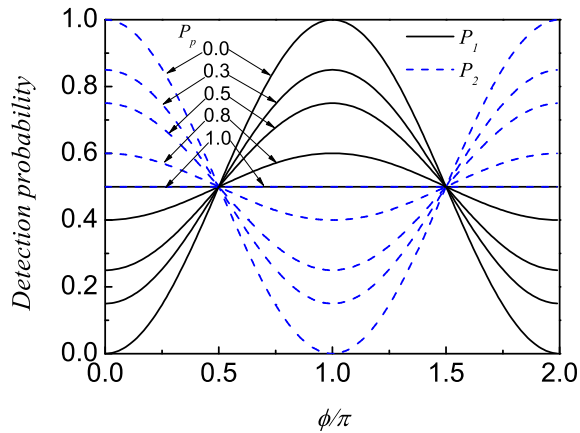}
\caption{The detection probabilities, $P_{1}$ and $P_{2}$, as functions of the phase $\phi$ at fixed
values of $P_{p}$. $P_{p}$ can be controlled by the BS$_2$ insertion instant of time that divides the passing sub-waves into different ratio between particle-like and wave-like parts. When $P_p=1.0$, BS$_2$ is not inserted, and no interference occur and the photon exhibits particle-like nature. When $P_p=0$, BS$_2$ is inserted before the sub-waves arrive at the exit point, and  full interference will occur, and the photon will show wave-like behavior. In between these two extremes, photon will exhibit partial particle-like nature and partial wave-like nature simultaneously as in the quantum delayed-choice case.   }\label{DPG}
\end{figure}
\end{center}
Apparently, $P_{1}^{w}=P_{2}^{w}=P_{w}/2$ and $P_{1}^{p}=P_{2}^{p}=P_{p}/2$,
where $P_{w}$ ($P_{p}$) is the total probability that will (will not) show interference (which is called wave (particle) nature in standard interpretation). Thus
\begin{equation}
P_{1}=\sin^{2}\frac{\phi}{2}+\frac{\cos\phi}{2}P_{p}
\end{equation}
and
\begin{equation}
P_{2}=\cos^{2}\frac{\phi}{2}-\frac{\cos\phi}{2}P_{p},
\end{equation}
and $P_{1}+P_{2}=1$. In the special case where $\phi=0$, BS$_{2}$ is
inserted when half of the two sub-waves
have exited the MZI, and this gives that $P_{1}=1/4$ and $P_{2}=3/4$.
$P_{1}$ and $P_{2}$ as functions of the phase $\phi$ at several fixed values of $P_{p}$ are shown
in Figure \ref{DPG}. It is seen that as $P_p$ changes from 0.0 to 1.0, the detection probabilities at the two arms change from a complete interference pattern to a flat line that exhibit no interference. Or in standard interpretation, the photon behavior changes from a wave to a particle. When the value of $P_p$ is fixed at a value between the two extremes, the probabilities are the incoherent superposition of a flat line and an interference pattern. In standard interpretation, a single photon exhibits wave nature and particle nature simultaneously.

This is equivalent to the quantum-delayed-choice experiment, where the controlled-insertion of the second beamsplitter serves as a controlled unitary gate that produces the superposed quantum state. The position of insertion gives the form of the unitary gate. At middle point insertion, the controlled gate is a Hadamard gate. This can also be explained in terms of the duality quantum computing framework in Ref.\cite{duality,duality2,duality3}, as in Ref.\cite{Roy}.

\vspace{0.5cm}
{\noindent \bf The EDC Experiment.}
We design and implement the EDC experiment in which the insertion of output beam splitter is decided
 at the end of the MZI when the photon is passing through the exit point. The experimental setup is shown in Fig. \ref{setup}.
\begin{widetext}
\begin{center}
\begin{figure}[!ht]
\includegraphics[width=15cm,angle=0]{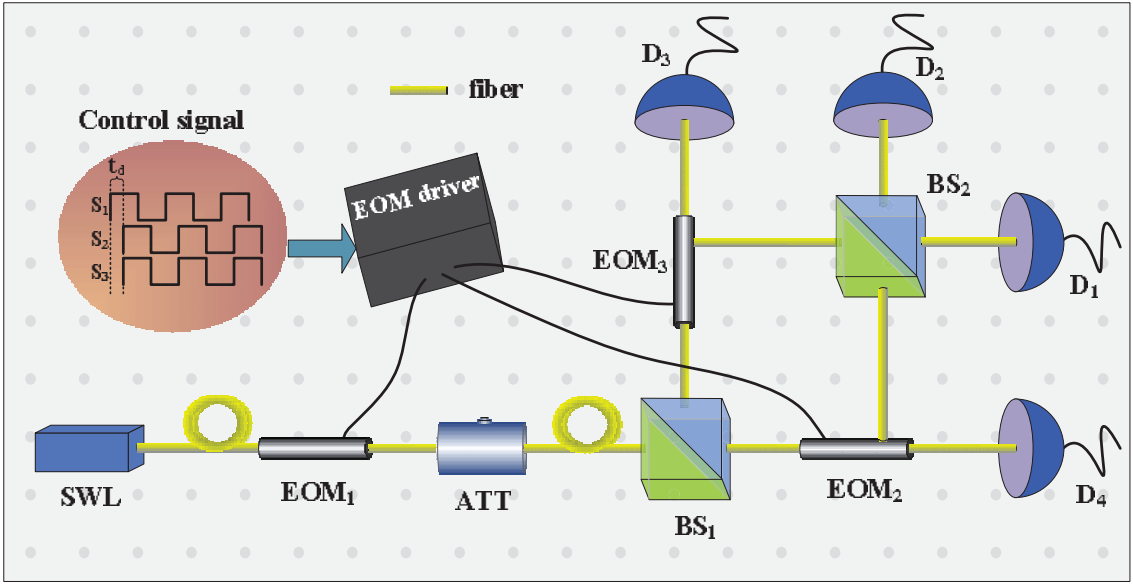}
\caption{Experimental realization of the EDC experiment.
SWL: Single-wavelength laser. EOM: Electro-optic modulator. ATT: Optical attenuator.
BS: Beam splitter. D: Single photon detector. Single photons are
produced by attenuating the pulses generated by  EOM$_1$ from a continuous light wave
emitted from a $780$ nm laser with a linewidth of $600$ kHz. The input and output beam
splitters are of 50:50 in transmission and reflection. The square waves TTL S$_{2}$ and S$_{3}$ signals apply
to the EOM$_{2}$ and EOM$_{3}$, respectively, which serves as a controller for insertion the second beamsplitter by guiding the sub-waves to different channels.
The control signals S$_{2}$ and S$_{3}$ are in-phase, and $t_{d}$ is the time difference
 between $S_1$ and S$_{2}$, S$_{3}$.}\label{setup}
\end{figure}
\end{center}
\end{widetext}
The experiment starts from a $780$ nm continuous-wave (CW) polarized laser (SWL) with a linewidth of $600$ kHz.
The first EOM$_{1}$ modulates and transforms the continuous light into pulse sequences, which then
are attenuated to the single-photon level  by using an attenuator. Then the pulses are sent
into the Mach-Zehnder interferometer, which are composed of two $50/50$ beam splitters and reflection mirrors. The input beamsplitter (BS$_{1}$) divides the wave function of a single photon into two spatially separated components of equal amplitude, and the output BS (BS$_{2}$) works
as a combiner of the two components.

The two arms of the MZI  are of equal lengths. The insertion of BS$_2$ is realized by using two additional modulators (EOM$_{2}$ and EOM$_{3}$), which are inserted in the two arms of the interferometer which are of equal distance from the input BS$_1$. The half-wave voltages of the three modulators are $V_{\pi}=91 \pm 1$ V. When the TTL signal is the "high" voltage level, the half-wave voltage applies to the EOM and the photon is transmitted, that is, the beamsplitter is lifted. Otherwise, the photon is reflected by the EOM, and the beamsplitter is inserted.

There are three TTL control signals with a repetition rate of $1$ MHz to determine whether or not
the half-wave voltages apply to the three modulators. EOM$_{1}$ is used to cut the continuous waves into fragmented pulses at the single photon level as explained earlier. The two modulators EOM$_{2}$ and EOM$_{3}$ are used to split the two sub-waves of the  single photon
into four sub-waves. When EOM$_{2}$ and EOM$_{3}$ are in the high voltage level, the two photon sub-waves are transmitted, and the MZI is open. The SUB-waves are directed to the detectors  D$_3$ and D$_4$ respectively, and  they show particle-like behavior. When the TTL are in the low voltage level, two of the sub-waves are reflected and pass through the output BS$_2$. Their paths are indistinguishable, and hence interfere with each other. The MZI interferometer is closed for them, hence show wave-like behavior in standard delayed-choice interpretation.

By maintaining the control signals S$_{2}$ and S$_{3}$ in-phase so that they act as a single one, and tune the time difference
$t_{d}$ between the signal S$_{1}$ and S$_{2}$. $t_{d}=0$ is the insertion time, namely, $t_d/(T/2)$ part of the sub-wave have transmitted, and move towards detectors $D_3$ and $D_4$, where $T/2$ is the length of the pulse.
The relative detection probability of $\text{D}_{3}$ is turned out to be
\begin{eqnarray}
R_{P}&=&\frac{_{out,1}^p\{\psi|\psi\}_{out,1}^p}
{_{out,1}^p\{\psi|\psi\}_{out,1}^p+_{out,2}^p\{\psi|\psi\}_{out,2}^p}\nonumber\\
&=&\frac{P_{1}^{p}} {P_{1}^p+P_{2}^{p}}=\frac{N_{3}}{N_{3}+N_{4}}\nonumber\\
&=&\frac{1}{2},
\end{eqnarray}
where $N_3$ and $N_4$ are the number of clocks registered by detectors $D_3$ and $D_4$ respectively. The result is independent of $t_d$, which is interpreted as exhibiting particle-like nature in standard interpretation. In REIN, this is naturally explained by the non-interfering sub-waves traveling through both arms simultaneously. The detection by either $D_3$ or $D_4$ is due to the measurement, which gives equal probabilities to each of the detectors.

On the other hand,  because of $\text{BS}_{2}$, the interference between the two sub-waves, $|\psi\}_{in,1}^w$ and $|\psi\}_{in,2}^w$, occurs. The two resulting sub-waves, $|\psi\}_{out,1}^w$ and $|\psi\}_{out,2}^w$, are then directed to detectors, $\text{D}_{1}$ and $\text{D}_{2}$. The relative detection probability of $\text{D}_{1}$ is evaluated as
\begin{eqnarray}
R_{W}&=& _{out,1}^w\{\psi|\psi\}_{out,1}^{w}\nonumber\\
&=&P^w_1(1-\cos\phi),\nonumber\\
&=&{N_1 \over N_{t} },\nonumber\\
\end{eqnarray}
where $N_1$ is the number of clicks registered by detectors $D_1$, and  $N_{t}=\sum^{4}_{i}N_{i}$. By choosing $\phi=0$, $R_{W}=0$ showing that destructive interference results in completely canceling each other in the output of $D_1$.

$P_{w}$ ($P_{p}$) is a probability that a single photon will (will not) show wave (particle) nature.
\begin{eqnarray}
P_{w}&=&_{out,1}^w\{\psi|\psi\}_{out,1}^w+_{out,2}^w\{\psi|\psi\}_{out,2}^w\nonumber\\
&=&2P_{1}^w\sin^2\phi/2+2P_1^w\cos^2\phi/2\nonumber\\
&=&2P_{1}^w=\frac{N_{1}+N_{2}}{N_{t}},\label{DW}
\end{eqnarray}
and
\begin{eqnarray}
P_{P}&=&_{out,1}^p\{\psi|\psi\}_{out,1}^p+_{out,2}^p\{\psi|\psi\}_{out,2}^p\nonumber\\
&=&2P_{1}^p=\frac{N_{3}+N_{4}}{N_{t}},\label{DP}
\end{eqnarray}
with $N_{t}=\sum^{4}_{i}N_{i}$ and $D_{W}+D_{P}=1$.

In our experiment, photon uniformly distributes in a pulse, yielding,
\begin{equation}
P_p=2t_d/T,
\end{equation}
and
\begin{equation}
P_w=1-P_p=1-2t_d/T.
\end{equation}
Both of $P_p$ and $P_w$ have linear relations with the delayed time $t_d$.

\begin{center}
\begin{figure}[!ht]
\includegraphics[width=7cm,angle=0]{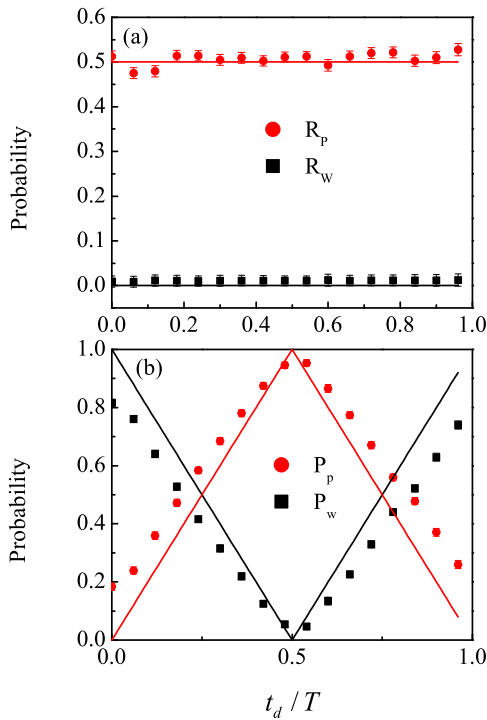}
\caption{Experimental results. (a) Black points represent ratio $R_w=N_1/(N_1+N_2)$ and red points are $R_p=N_3/(N_3+N_4)$, which represents  wave-like behavior and  particle-like behavior respectively in standard interprettaion; (b) The total probability $P_w$ of interfering photon (black dots) and the that, $P_p$, of non-interfering photon (red dots) .}\label{data}
\end{figure}
\end{center}
The experimental results are shown in Fig. \ref{data}.  It is seen that
the wave function of a single photon is divided into four parts and detected by four
detectors, respectively. If the output BS is present, we will observe the interference fringes
with a tunable phase difference between the two paths which the single photon sub-waves travel through.
When the two arms of the interferometer are of equal length, the two paths are fully
recombined by the output BS and perfectly indistinguishable.
We register, with probability $1$, a click in only one of the two detectors
(D$_{1}$ and D$_{2}$) placed on the output ports of the interferometer.

If the output BS is absent,
each of the detector has 50\% probability to register a click. In the standard interpretation, this is interpreted as the photon having
 particle-like behavior, and the photon travels through a single path to each one of the detector. In the REIN view, this is interpreted in a unified way just as the closed setting case. The only difference is whether or not BS$_2$ exists. Before the exit point, sub-waves travel in both arms.  Without BS$_2$, sub-waves travel without  interference, and with BS$_2$ sub-waves interferes that may lead the photon wave to go one detector completely.

 As seen in Figure \ref{data}(a), the black points $R_{W}=N_{1}/(N_{1}+N_{2})$ shows the wave-like behavior,
and the red ones represent $R_{P}=N_{3}/(N_{3}+N_{4})$  shows the particle-like behavior. $P_{W}$ gives the percentage
of the component of the the single photon wave function showing wave-like
behavior and $D_{P}$ gives that of the component showing
particle-like behavior.  It allows the ratios $D_{W}$ and $D_{P}$
to vary between $0$ and $1$ when the time delay $t_{d}$ varies between $0$ and $T$, where $T$ is the
period of the control signal, where $T/2$ are in high voltage level and $T/2$ are in low voltage level. The wave function of the single photon distributes with uniform
intensity along the propagation direction in virtue of the rectangular control signals with $50\%$ duty cycle. Because the frequency of the control
signal, $f=1/T$, is larger than the laser linewidth of $600$ kHz, the coherence length of the light modulated
by EOM$_{1}$ approaches to that of the pulse and the length of the the single photon wave
function along the propagation direction could be considered as that of the pulse $L={Tc}/(2n)$
with the light speed $c$ and the effective refractive index $n$. Hence, the two quantities $D_{W}$ and
$D_{P}$  change linearly with the time-delay $t_d$ as shown in Figure \ref{data}(b).

\vspace{0.5cm}
{\noindent \bf Discussion} \ \

In this work, we have presented the realistic interpretation of quantum mechanics, the REIN. In REIN, the wave function, or wave are the real existence of quantum object. It is not merely a mathematical description. Like classical wave,  quantum wave can be divided into sub-waves, and the sub-waves can be recombined. When they are measured, they collapse and show the particle-like nature. The essential difference between quantum wave and classical wave is that quantum wave collapses in totality, namely the whole of the quantum wave, whatever scattered in space, will collapse into a single point instantly. Apart from this, quantum wave can be almost viewed in the same manner as classical wave.

In the REIN view, in the MZI device, the photon sub-waves travel through both arms. The simultaneous travel of a photon through the two arms is easy to comprehend and understand in REIN: photon is no longer a ball-like particle, it is an extended, and even separated stuff distributed in space, the quantum wave, or quantum sub-waves. The sub-waves travel simultaneously through the two arms. Each sub-wave contains the full attributes of the quantum object: when measured, it collapses with certain probability to exhibit the full properties of the quantum object, such as spins, masses and so on.

In the REIN view, the wave-like nature or particle-like nature in the standard interpretation of delayed-choice MZI, is simply the interference or non-interference of the sub-waves of the single photons. In the REIN view, the photons are all sub-waves before they are detected. When they are are detected, they collapse and cause a click in the detector which is viewed as a particle.

The REIN view has been exploited in the duality quantum computer \cite{duality}. The duality quantum computer uses the superpositions of quantum sub-waves, and hence allows the linear combinations of unitary operators as generalized quantum gates. The mathematical expressions have been constructed and developed \cite{gudder1,duality4,duality3p,duality5,duality6}. Recently, it has been found that linear combinations of unitary operators are superior in simulating Hamiltonian systems over traditional formalism of products of unitary operators \cite{childs}.

The REIN idea is more detailed demonstrated by an encounter-delayed-choice experiment proposed in this work. By inserting a beam-splitter during the encounter of two sub-waves, one is able to let part of the sub-waves to interfere and the other part not to interfere, hence exhibiting the so-called wave-like nature and particle-like nature simultaneously as in the quantum delayed-choice experiment. We have experimentally demonstrated the EDC proposal, and the experiment results support the REIN idea.

{\bf Acknowledgments} This work is supported by the National Natural Science
Foundation of China under Grants No.11474181, the National Basic Research Program of China under Grant No.
2011CB9216002 and the Open Research Fund Program of the State Key Laboratory
of Low-Dimensional Quantum Physics, Tsinghua University.

{\bf Author contributions} GLL conceived the realistic interpretation idea, the encountered-delayed-choice experimental proposal, and supervised the whole project. WQ, JLL and GLL setup the experimental apparatus, WQ, JLL, ZY and GLL performed the experiment, WQ and GLL analyzed the data. GLL and WQ wrote the paper.

{\bf Competing financial interests} The authors declare no competing
financial interests.

{\bf Correspondence} and requests for materials should be addressed
to G.-L.L. (gllong@mail.tsinghua.edu.cn).

\end{document}